\documentclass{jfm}
\usepackage{graphicx}
\usepackage{epstopdf, epsfig}
\usepackage{amsmath}
\usepackage[mathlines]{lineno}

\shorttitle{Signatures of fluid-fluid displacement in porous media}
\shortauthor{B.K. Primkulov et al.}

\title{Signatures of fluid-fluid displacement in porous media: wettability, patterns, and pressures}

\author{Bauyrzhan K. Primkulov\aff{1},
    Amir A. Pahlavan\aff{2},
    Xiaojing Fu\aff{3},
    Benzhong Zhao\aff{4},
    Christopher W. MacMinn\aff{5},
    \and Ruben Juanes\aff{1}\corresp{\email{juanes@mit.edu}}}

\affiliation{\aff{1}Massachusetts Institute of Technology, 77 Massachusetts Avenue, Cambridge, MA, USA
\aff{2}Princeton University, Olden St., Princeton, NJ, USA
\aff{3}University of California at Berkeley, McCone Hall, Berkeley, CA, USA
\aff{4}McMaster University, 1280 Main Street West, Hamilton, Canada
\aff{5}University of Oxford, Parks Road, Oxford, United Kingdom
}

\begin{document}

\maketitle

\begin{abstract}
We develop a novel ``moving capacitor'' dynamic network model to simulate immiscible fluid-fluid displacement in porous media. Traditional network models approximate the pore geometry as a network of fixed resistors, directly analogous to an electrical circuit. Our model additionally captures the motion of individual fluid-fluid interfaces through the pore geometry by completing this analogy, representing interfaces as a set of moving capacitors. By incorporating pore-scale invasion events, the model reproduces, for the first time, both the displacement pattern and the injection pressure signal under a wide range of capillary numbers and substrate wettabilities. We show that at high capillary numbers the invading patterns advance symmetrically through viscous fingers. In contrast, at low capillary numbers the flow is governed by the wettability-dependent fluid-fluid interactions with the pore structure. The signature of the transition between the two regimes manifests itself in the fluctuations of the injection pressure signal.
\end{abstract}


\section{Introduction}\label{sec:intro}
A beautiful array of flow patterns arises when a low-viscosity fluid displaces a more-viscous fluid in a porous medium. The problem has been extensively examined through laboratory experiments, as well as numerical simulations and theoretical models \citep{Saffman1958,Bensimon1986,Homsy1987,Paterson1981,Tryggvason1983,Nittmann1985,Kadanoff1985,Arneodo1989,Li2009,Bischofberger2015,Chen1985,Maloy1985,Chen1987,Fernandez1990}. The dynamics of such displacement can be characterized by two dimensionless groups: the ratio of viscous to capillary forces, or the capillary number ($\text{Ca}$), and the ratio of defending to invading fluid viscosities, or viscosity contrast ($M$). For high $\text{Ca}$, the resulting displacement patterns are reminiscent of diffusion limited aggregation \citep{Witten1981,Daccord1986,Meakin1989,Niemeyer1984,Conti2010}. For low $\text{Ca}$, the displacement dynamics becomes more intricate, and the emerging patterns display a strong dependence on the pore geometry \citep{Lenormand1985,Lenormand1983,Lenormand1988,Fernandez1991,Maloy1992,Furuberg1996,Ferer2004,Toussaint2005,Holtzman2012} and the wettability of the medium, that is, the chemical affinity of the solid for each fluid \citep{Stokes1986,Trojer2015,Zhao2016,Odier2017}. In particular, an intermittent injection pressure signal emerges in the limit of low $\text{Ca}$ \citep{Furuberg1996,Maloy1992}. Given that in most practical applications visualization of the flow in porous media is not possible, the pressure signal is often the only source of information. Surprisingly, no modeling approach to date has been able to capture the injection pressure signal across different $\text{Ca}$ and pore wettabilities. Here, we develop a new pore-network model that fills this gap, and we use it to explore the transition from viscous-dominated to capillary-dominated flow regimes by examining the connections among fluid morphology and pressure signal.

Pore network models of flow in porous media can be broadly classified into two groups: quasi-static and dynamic models \citep{Blunt2001,Meakin2009,Joekar-Niasar2012}. Quasi-static models neglect viscous effects and advance the invading fluid through either invasion-percolation \citep{Chandler1982, Lenormand1988} or event-based algorithms \citep{Cieplak1990,Cieplak1988}. Although a quasi-static approach can be effective in reproducing experimental invasion patterns at low $\text{Ca}$ \citep{primkulov18-prf}, it is unable to capture the temporal evolution of the injection pressure signal. Dynamic network models approximate the flow channels with a network of interconnected capillary tubes. Viscous pressure drops are calculated by assuming fully developed viscous flow within each tube. Local capillary pressures within the network are calculated from either the interface position within pore throats \citep{Aker1998,Gjennestad2018} or through mass balance of the two phases in pore bodies \citep{Al-Gharbi2005,JOEKAR-NIASAR2010}. 
Another notable class of models is invasion-percolation in a gradient: a percolation model designed to incorporate buoyancy forces \citep{wilkinson84,birovljevfuruberg91, frettefeder92, meakinfeder92}, and then extended to model (linear) pressure gradients \citep{yortsosxu97}. None of the invasion-percolation in a gradient studies, however, incorporate any notion of wettability (they all deal exclusively with strong drainage), pore-scale dynamics, or capillary-number-dependent pressure fluctuations.

In fact, most existing pore-network models, both quasi-static and dynamic, are limited to strong drainage (or injection of non-wetting fluid) and do not include wettability-induced cooperative pore filling \citep{JOEKAR-NIASAR2010,Aker1998,Al-Gharbi2005,Holtzman2010}. The only dynamic pore network model to date that includes cooperative pore filling events \citep{Holtzman2015} does so by combining pore-level invasion events of \citet{Cieplak1988,Cieplak1990} with viscous relaxation through the pore-network. This viscous-relaxation assumption is at odds with the physics of interface motion in the capillary-dominated regime and, as a result, this model is unable to capture the injection pressure signal observed experimentally in the limit of intermediate and low $\text{Ca}$ \citep{Zhao2016,Furuberg1996,Maloy1992}. We present in \S\ref{sec:model} a consistent framework that combines viscous, capillary, and wettability effects in a single dynamic network model that builds a direct analogy between local fluid-fluid interfaces and electric capacitors. Our model reproduces, quantitatively, the fluid-fluid displacement patterns for a wide range of $\text{Ca}$ and wettabilities (\S\ref{sec:patterns}), and points to a surprising and heretofore unrecognized transition in the pressure fluctuations between the low and high Ca flow regimes (\S\ref{sec:pressure}).

\section{Moving Capacitor Model}\label{sec:model}

\begin{figure}
\centering
 \includegraphics[width=.95\linewidth]{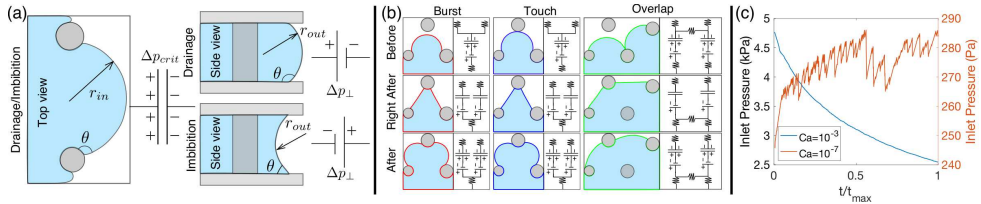}
 \caption{\label{fig:fig_1}(a)~Schematic diagram of in-plane and out-of-plane curvatures within the flow cell. Out-of-plane curvature represents the overall affinity of the porous medium to the invading fluid. It is determined by $\theta$ and is analogous to a battery. In-plane curvature changes as the local interface evolves while pinned to a pore throat, and it is analogous to a capacitor. (b)~Evolution of burst, touch, and overlap events. (c)~Temporal profiles of the injection pressure bear close resemblance to similar experiments in the drainage regime at low (orange) and high (blue) $\text{Ca}$ \citep{Furuberg1996,Zhao2016}.}
\end{figure}

Consider a moving fluid-fluid interface in a micromodel (FIG \ref{fig:fig_1}a). Neglecting dynamic-contact-angle effects \citep{Hoffman1975} for simplicity, the shape of the meniscus between posts is uniquely defined by the combination of Laplace pressure and substrate wettability defined through a contact angle $\theta$ at which the interface meets post surfaces \citep{Cieplak1988,Cieplak1990}. As the interface advances, the Laplace pressure increases until the interface encounters a \textit{burst}, \textit{touch} or \textit{overlap} event, as defined by \citet{Cieplak1988,Cieplak1990}. The \textit{burst} event is equivalent to a Haines jump \citep{Haines1930,Berg2013}, while the \textit{touch} and \textit{overlap} events take place when the local interface either touches the nearest opposing post or coalesces with a neighboring interface respectively [FIG. \ref{fig:fig_1}(b)]. If the interface becomes unstable due to \textit{burst} or \textit{touch}, a single pore is invaded and two new interfaces appear. In the case of an \textit{overlap} event, two (in some cases more) pores are filled simultaneously. 
These pore-level events are an integral part of the model and, indeed, this sensitivity is what permits capturing wettability effects within the model. The events evolve differently at different wettabilities---burst events are most frequent in drainage, while touch and overlap are most frequent in imbibition (or injection of wetting fluid) \citep{Cieplak1990,primkulov18-prf}.

We can explicitly calculate the critical Laplace pressure $\Delta{p_{\text{crit}}}$ corresponding to all events from the values of the contact angle, radii and coordinates of the posts \citep{primkulov18-prf}, and thus can use the analogy between electric capacitors and fluid-fluid interfaces in constructing our network model. A capacitor represents the pinning of the fluid--fluid interface at a pore throat, and is active in both drainage and imbibition: the interface moves only when a local depinning threshold ($\Delta{p_\text{crit}}$) is reached, and the fluid front moves to restart the pinning--depinning cycle from zero in-plane curvature [Fig.~1(b)]. This progression of the in-plane curvature in our model was motivated by the work of Cieplak and Robbins \citep{Cieplak1988,Cieplak1990} [see also \citep{Rabbani2018}] and experiments on the progression of the in-plane curvature between the Hele-Shaw cell posts \citep{Jung2016,Lee2017}. This is what allows capturing pressure fluctuations in the limit of low $\text{Ca}$ [Fig.~1(c)]. The battery analogy represents the overall affinity of the porous medium to the invading fluid, set by the out-of-plane curvature at the fluid front. The out-of-plane curvature is fixed throughout a single simulation, and determined by the value of the contact angle (given the constant gap between the flow-cell plates): it is positive in drainage and negative in imbibition [Fig.~1(a)]. To complete the analogy between an electric circuit and a pore network, one can think of a network of resistors being responsible for viscous effects, capacitors and batteries responsible for capillary effects, and local rules for circuit rearrangements responsible for wettability effects [FIG. \ref{fig:fig_1}(b)].

Therefore, the pressure drop across an edge of the network containing a fluid-fluid interface has three components: (i) pressure drop due to viscous dissipation, (ii) Laplace pressure drop due to in-plane curvature of the interface, and (iii) Laplace pressure drop due to out-of-plane curvature of the interface. We calculate the viscous pressure drop assuming Poiseuille flow in a capillary tube, which is analogous to the potential drop across a resistor. The out-of-plane component of the Laplace pressure can be expressed as either a positive or negative pressure jump ($\Delta{p_{\perp}}=\displaystyle -\frac{2\gamma \cos\theta}{h}$, where $\gamma$ is the interfacial tension, and $h$ is the cell height) depending on the substrate wettability; this is analogous to a battery in an electric circuit. The Laplace pressure due to in-plane curvature of the interface is analogous to a capacitor which allows flow until it reaches the critical pressure ($\Delta{p_{\text{crit}}} = \text{min}\{p_{\text{burst}},p_{\text{touch}},p_{\text{overlap}}\}$). Since we can calculate $\Delta{p_{\text{crit}}}$ for all edges at the invading fluid front, we use a linear estimate of the in-plane Laplace pressure drops within our network ($\Phi(t) \Delta{p_{\text{crit}}}$), where $\Phi(t)$ stands for the filling ratio of a given throat. When $\Phi(t) \rightarrow 0$, the in-plane Laplace pressure is negligible. When $\Phi(t) \rightarrow 1$, the throat is nearly full and has a critical in-plane Laplace pressure $\Delta{p_{\text{crit}}}$. This analogy between local interfaces and capacitors allows us to incorporate local changes in Laplace pressure due to filling of pore throats. Once a node in the network reaches its maximal potential, which coincides with its filling capacity, it becomes unstable and the interface advances. We assume that the in-plane and out-of-plane Laplace pressures are decoupled, and this is done to maintain the simplicity of the overall model. With this assumption, one can run the model for either $\frac{h}{a} \gg 1$ or $\frac{h}{a} \ll 1$, where these conditions would result in negligible or dominant contributions of the out-of-plane curvature in the model, respectively.

\begin{figure}
 \centering
 \includegraphics[width=.9\linewidth]{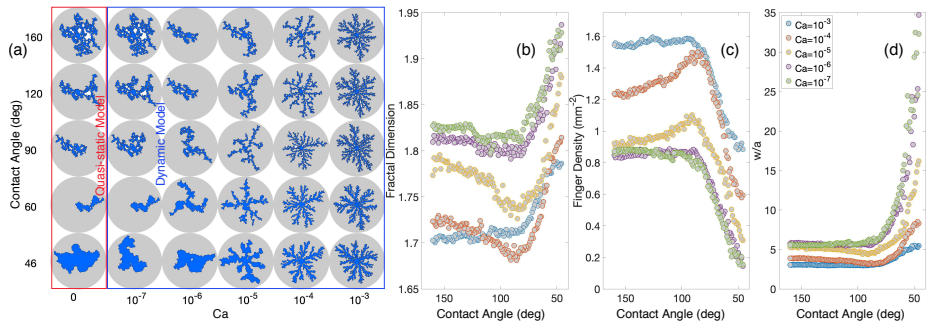}
 \caption{\label{fig:fig_2} (a)~Phase diagram of the invading fluid morphology at breakthrough; (b)~Fractal dimension, computed by means of the box-counting method; (c)~Number of fingers per unit area of injected fluid, which exhibits a maximum near $\theta=90^\circ$; (d)~Normalized finger width ($w/a$) at different $\text{Ca}$ and wettabilities measured at breakthrough. Finger width increases as the posts become more wetting to the invading fluid.}
\end{figure}

The topology of the pore network is captured through the incidence matrix $A$ by examining the adjacency of the pores \citep{Strang2007}. We number all pores and adopt the convention that pore connections are oriented in the direction of increasing pore numbers. Rows of $A$ represent edges, and columns of $A$ represent nodes of the network. We also make use of the diagonal conductance matrix $C$, whose elements are hydraulic conductivities of the network edges. The elements of this matrix can be calculated as $c = \frac{\pi r^4}{8 \mu L}$, assuming fully developed Hagen-Poiseuille flow through a rectangular tube with hydraulic radius $r$ and length $L$, where $\mu$ is the effective viscosity of the fluid in the channel.

The pressure difference across the network edges can be calculated as $e=b-Ap$, where $b$ and $p$ stand for pressure change due to out-of-plane contribution to Laplace pressure (batteries) and node pressures, respectively. The network flow rates can be calculated from this pressure difference as $q=Ce$. At the same time, flow rates must obey mass conservation, $A^Tq=f$, where $f$ stands for flow sources at the nodes. After eliminating $e$, the flow through the network without the in-plane contribution to Laplace pressure (capacitors) is obtained through the following system of equations:

\begin{align}
    q &= C(b-Ap), \label{eqn:single_flow_ungrounded1} \\
    A^{T}q &= f. \label{eqn:single_flow_ungrounded2}    
\end{align}
We set constant flow boundary conditions at the inlet pores (at the center of the flow cell) and constant pressure boundary conditions at the outlet pores (at the edges of the flow cell). We note that $Ap$ can be decomposed into components of nodes with prescribed pressure and all other nodes ($Ap = A_{\text{outer}}p_{\text{outer}} + \tilde{A}\tilde{p}$), and therefore Eqs.~\eqref{eqn:single_flow_ungrounded1}-\eqref{eqn:single_flow_ungrounded2} transform to:
\begin{eqnarray}
\begin{bmatrix}
    C^{-1} & \tilde{A} \\
    \tilde{A^T} & 0 
\end{bmatrix}
\begin{bmatrix}
    q \\
    \tilde{p} 
\end{bmatrix}
=
\begin{bmatrix}
    b-A_{\text{outer}}p_{\text{outer}} \\
    \tilde{f} 
\end{bmatrix}
=
\begin{bmatrix}
    \tilde{b} \\
    \tilde{f} 
\end{bmatrix}.
\label{eqn:single_phase_final}
\end{eqnarray}
The solution to \eqref{eqn:single_phase_final} provides values of both edge flow rates and node pressures for given boundary conditions.

Finally, we incorporate the pressure drop due to in-plane Laplace pressure (capacitors) within the network. Taking into account the direction of the edges (an array $d(t)$ consisting of $1$ and $-1$), the total pressure drop across the network edges can be written as $e = \tilde{b} - \tilde{A}\tilde{p} - d(t)\Phi(t)\Delta{p_{\text{crit}}}$. In other words, the in-plane Laplace pressure is the product of the filling ratio and the critical pressure from the quasi-static model \citep{primkulov18-prf}. Therefore, the equations governing two-phase flow through the network can be written as:
\begin{eqnarray}
\begin{bmatrix}
    C^{-1}(t) & \tilde{A} \\
    \tilde{A^T} & 0 
\end{bmatrix}
\begin{bmatrix}
    q(t) \\
    \tilde{p}(t) 
\end{bmatrix}
=
\begin{bmatrix}
    \tilde{b}- d(t)\Phi(t)\Delta{p_{\text{crit}}}\\
    \tilde{f} 
\end{bmatrix}.
\label{eqn:two_phase_final}
\end{eqnarray}

We now discuss the mechanics of the time-stepping in our two-phase flow model. After we initialize the interface locations within the network, we use an adaptive forward Euler time stepping to update the filling ratios of the network edges at the interface $\Phi(t)$. We ensure that only a fraction of the edge total volume at the interface flows within the time-step \citep{Aker1998}. After every time-step, we use $\Phi(t)$ to update the conductance matrix $C(t)$ and resolve the flow through Eq.~\eqref{eqn:two_phase_final} with updated pressure drops across the fluid-fluid front.

In the spirit of the fundamental contributions from Cieplak and Robbins \citep{Cieplak1988,Cieplak1990}, our model takes the form of an arrangement of cylindrical posts confined between the plates of a Hele-Shaw cell. The approach is simple enough to lead to universal findings, yet sufficiently complex to have direct relevance to microfluidic geometries, as well as engineered and natural porous media---much like Lenormand's phase diagram \citep{Lenormand1988}. By doing so, we demonstrate the ability to reproduce physics---in particular, pressure fluctuations under a wide range of wetting conditions---which, until now, were inaccessible to pore-network modeling. A limitation of the model presented here is that it does not extend to contact angles below 45$^\circ$, where the wetting fluid preferentially wets the corners of the pore-geometry at low $\text{Ca}$ and forms film flow at high $\text{Ca}$ \citep{Zhao2016, Odier2017}.

\section{Invasion Patterns}\label{sec:patterns}
We simulate immiscible fluid-fluid displacement by setting a constant injection rate at the center of the flow cell and zero pressure at the outlets. The invading and defending fluid viscosities are set to $8.9 \times 10^{-4} ~ \text{Pa} \cdot \text{s}$ and $0.34 ~ \text{Pa} \cdot \text{s}$ respectively. The post height $h$ is $100\mu$m, and interfacial tension $\gamma$ is set to $13 \times 10^{-3}$ N/m. These parameters as well as the pore geometry are chosen to mimic the experiments of \citet{Zhao2016}. The flow cell has an outer diameter of $30~$cm. We perform simulations for wetting conditions from strong drainage ($\theta=160^\circ$) to weak imbibition ($\theta=46^\circ$). FIG. \ref{fig:fig_1}(c) shows the pressure profiles for $\theta=160^\circ$ at $\text{Ca} \in \{10^{-3},10^{-7}\}$, respectively. In the limit of high $\text{Ca}$, the more-viscous defending fluid sustains substantial spatial pressure gradients, and the injection pressure gradually drops as more of the defending fluid is displaced \citep{Zhao2016}. In contrast, in the limit of low $\text{Ca}$, the pressure field is virtually uniform in each fluid, and the injection pressure exhibits intermittent fluctuations typical of slow capillary-dominated drainage \citep{Knudsen2002,Aker1998a,Maloy1992,Moebius2012}.

The morphology of the invading fluid at breakthrough can be analyzed by means of a binary-image representation of the invasion patterns \citep{Cieplak1990,Cieplak1988,primkulov18-prf} [FIG. \ref{fig:fig_2}(a)]. We estimate the width and number of fingers in the invading fluid pattern following the protocol outlined in \citep{Cieplak1988,Cieplak1990} and modified in \citep{primkulov18-prf}. The binary image is sliced horizontally and vertically, with each slice containing clusters of invading fluid pixels. We calculate the finger width as the mean size of these clusters. FIG. \ref{fig:fig_2}(d) shows that the finger width, normalized by the typical pore size, increases as $\theta \rightarrow 46^\circ$ for all $\text{Ca}$, which is in agreement with experimental observations \citep{Stokes1986,Trojer2015,Zhao2016}. While FIG. \ref{fig:fig_2}(a) demonstrates that the number of fingers increases with $\text{Ca}$ \citep{Lenormand1988,Fernandez1990,Zhao2016}, we observe an unexpected behavior [FIG. \ref{fig:fig_2}(b)]: the finger density changes with the substrate wettability, and exhibits a maximum around $\theta=90^\circ$. This effect is most pronounced for $10^{-6} < \text{Ca} < 10^{-3}$ (when viscous and capillary effects are comparable).

We explain the peak in the viscous finger density at $\theta \approx 90^\circ$ in FIG. \ref{fig:fig_2}(b) by considering in-plane and out-of-plane contributions to the Laplace pressure. At a fixed $\text{Ca}$, the ratio of viscous and capillary forces in the micromodel changes as a function of substrate wettability. The capillary forces have out-of-plane contributions, which are nominally equal to zero when $\theta=90^\circ$, so the ratio of viscous and capillary forces increases as $\theta$ changes from $160^\circ$ to $90^\circ$ at fixed $\text{Ca}$. In addition, when $\theta$ changes from $90^\circ$ to $46^\circ$, the cooperative pore filling mechanisms become dominant and widen the largest fingers, which in turn consume the smaller ones and reduce the number of fingers. The combination of these two effects results in the local maximum in the number of viscous fingers around $\theta \approx 90^\circ$ across different $\text{Ca}$ [FIG. \ref{fig:fig_2}(b)]. 

For a contact angle~$\theta$ near $160^\circ$ (strong drainage) and high values of $\text{Ca}$ ($10^{-3}$ and $10^{-4}$), the invading fluid front advances through viscous fingers with fractal dimension close to~$1.71$, typical of DLA-type morphology \citep{Witten1981}. As $\text{Ca}$ is reduced to a low value ($10^{-7}$), the fractal dimension increases to about~$1.82$, characteristic of invasion-percolation \citep{Wilkinson1983} [FIG.~\ref{fig:fig_2}(b)]. This increasing trend in fractal dimension is consistent with the decrease in finger density [FIG.~\ref{fig:fig_2}(c)] and the increase in finger width [FIG.~\ref{fig:fig_2}(d)].
    
As the contact angle approaches $46^\circ$, cooperative pore filling becomes the dominant flow mechanism at all values of~$\text{Ca}$. This flow regime results in the compact displacement of the defending fluid, and thus the fractal dimension increases, approaching a value of~$2$ at low $\text{Ca}$, indicative of stable displacement.

\section{Pressure Signature}\label{sec:pressure}

\begin{figure}
 \centering
 \includegraphics[width=.95\linewidth]{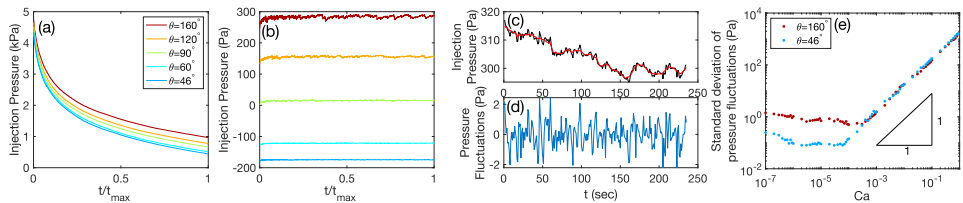}
 \caption{\label{fig:fig_3} (a)-(b) Temporal evolution of the injection pressure at $\text{Ca}=10^{-3}$ and $\text{Ca}=10^{-7}$ respectively. At high $\text{Ca}$, the injection pressure decreases as the viscous fingers approach the outer boundary of the flow cell. At low $\text{Ca}$, the injection pressure is dominated by Laplace pressure fluctuations at the interface. We use wavelet decomposition \citep{Cai2002,Sygouni2006,Sygouni2007} to split the pressure signal ($\text{Ca}=10^{-5}$ and $\theta=160^\circ$ here) into its (c) global trend and (d) cyclic component. (e) The standard deviation of the pressure fluctuations point at two different regimes. At low $\text{Ca}$, pressure fluctuations are dominated by stick-slip changes in Laplace pressure. At high $\text{Ca}$, pressure fluctuations are dominated by changes in the effective hydraulic conductance of dominant flow channels.}
\end{figure}

The fundamental difference in the fluid-fluid displacement process between low and high $\text{Ca}$ is reflected in the temporal injection-pressure signals [FIG. \ref{fig:fig_3}]. When the capillary number is relatively high ($\text{Ca}=10^{-3}$), viscous forces dominate, and the injection pressure decreases with time for all substrate wettabilities \citep{Zhao2016} [FIG. \ref{fig:fig_3}(a)]. Here, most of the pressure drop takes place in the more-viscous defending fluid. Consequently, as more of the defending fluid is displaced, the pressure required to maintain the prescribed injection flow rate decreases. In contrast, at $\text{Ca}=10^{-7}$, viscous dissipation is negligible, and the injection pressure is determined by the sum of outlet and Laplace pressures. As a result, the injection pressure fluctuates in a stick-slip manner around a mean value [FIG. \ref{fig:fig_3}(b)], as has been documented in slow drainage experiments \citep{Maloy1992,Furuberg1996,Moebius2012}. The pressure signals in FIG. \ref{fig:fig_3}(b) highlight the roles that in-plane and out-of-plane curvatures play in our model. Out-of-plane curvature plays the role of batteries, and thus provides additional resistance/drive (in drainage/imbibition, respectively) to the flow at the interface. The magnitude of the pressure drop/rise at the batteries is a function of wettability, which explains why the mean value of the injection pressure signal also varies with wettability [FIG. \ref{fig:fig_3}(b)]. The in-plane curvature plays the role of capacitors. As the invading fluid is injected, the in-plane component of Laplace pressure grows at the interface until the meniscus near the pore with lowest critical entry pressure becomes unstable due to \textit{burst}, \textit{touch} or \textit{overlap}. This results in the rapid advance of the local interface, which pressurizes the defending fluid ahead. This overpressure then dissipates (see video S1 in supplemental materials). The critical pressures of \textit{touch} and \textit{overlap} are always smaller than the critical pressures of \textit{burst} events \citep{primkulov18-prf,Cieplak1990,Cieplak1988}, so the magnitude of the pressure fluctuations decreases as the substrate becomes more wetting to the invading fluid [FIG. \ref{fig:fig_3}(b)].

To gain further insight into the difference in the pressure signature between low and high $\text{Ca}$, we decompose the injection pressure signal into its global trend and fluctuating components with Block James-Stein wavelet decomposition \citep{Cai2002} (see FIG. \ref{fig:fig_3}c-d). We compute the standard deviation of the fluctuating component of the pressure signal for both drainage and imbibition conditions ($\theta = 160^\circ$ and $46^\circ$, respectively) for a wide range of $\text{Ca}$, and find that it exhibits two distinct regimes [FIG. \ref{fig:fig_3}(e)]. At low $\text{Ca}$, pressure fluctuations are controlled by the stick-slip-type changes in local Laplace pressures. In contrast, at high $\text{Ca}$, pressure fluctuations are controlled by changes in the effective hydraulic conductance of the dominant flow channels. In the limit of high $\text{Ca}$, the Laplace pressure drop is negligible in comparison with the viscous pressure gradient, but the dominant flow channels are rearranged slightly as the fingers grow (see video S2 in supplementary materials). Since the pore geometry has a heterogeneous distribution of throat sizes, shifts in the dominant flow channels result in viscosity-driven pressure fluctuations at high $\text{Ca}$.

\begin{figure}
    \centering
    \includegraphics[width=.4\linewidth]{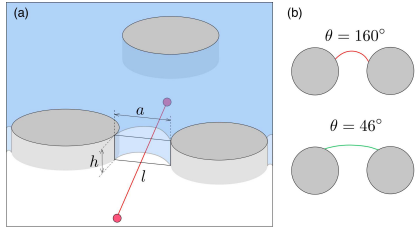}
    \caption{(a) Pore-scale perspective for the scaling of pressure fluctuations. The diagram shows a typical pore being invaded. The characteristic distance between the pore centers is $l$ (red line), the post height is $h$, and a characteristic throat size is $a$. (b) Typical configurations of the fluid-fluid interface in drainage and imbibition. Burst events are prevalent in drainage and the typical radius of out-of-plane curvature is of order $a$. Overlap events are prevalent in imbibition and the typical radius of out-of-plane curvature is an order of magnitude greater than $a$. }
    \label{fig:scaling}
\end{figure}

Scaling arguments support the findings from the model simulations. Let us take a pore-scale perspective (see Fig.~\ref{fig:scaling}). Invading a single pore involves overcoming a capillary pressure and pushing defending fluid out through a throat of width $a$ and height $h$ at a speed proportional to the injection rate. The capillary pressure is $p_\text{cap} \approx \gamma (\frac{1}{h}+\frac{1}{a f(\theta)})$, where $f(\theta)$ is a wettability-dependent function that takes a value $\sim 1$ near drainage and $\sim 10$ near strong imbibition [Fig.~\ref{fig:scaling}(b)]. Taking variations of $p_\text{cap}$ with $a$ yields
\begin{equation}
    \delta{p_\text{cap}} \sim \frac{\gamma}{a^2 f(\theta)} \delta{a}.
    \label{eqn:cap_fluc}
\end{equation}
The characteristic flow velocity through a typical throat is $u = \frac{k(a,h)}{\mu} \frac{p_\text{visc}}{l}$, where $k(a,h)=R_h^2/8$ is the rectangular channel permeability and $R_h=\frac{ah}{2(a+h)}$ the hydraulic radius. Thus the viscous pressure is $p_\text{visc} \sim \frac{32 (a+h)^2}{a^2 h^2} \mu u l = \frac{32 \mu u l}{h^2} (1+h/a)^2$. Taking variations of $p_\text{visc}$ with $a$ yields
\begin{equation}
    \delta{p_\text{visc}} \sim \frac{64 \mu u l}{h^2} (1+h/a) \frac{h}{a^2} \delta{a} = \frac{64 \mu u l}{h a^2} (1+h/a) \delta{a}.
    \label{eqn:visc_fluc}
\end{equation}
The magnitude of the total characteristic pressure fluctuation is $\delta{p_\text{cap}} + \delta{p_\text{visc}}$, and its two components are comparable when $\frac{\delta{p_\text{visc}}}{\delta{p_\text{cap}}} \sim 1$. Using equations (\ref{eqn:cap_fluc}) and (\ref{eqn:visc_fluc}),
\begin{equation}
    \frac{\delta{p_\text{visc}}}{\delta{p_\text{cap}}} \sim \frac{64 \mu u l}{h a^2} (1+h/a) \frac{a^2 f(\theta)}{\gamma} = \text{Ca} f(\theta) 64 \frac{l}{h} (1+h/a) \sim 1,
\end{equation}
which implies a crossover $\text{Ca}$,
\begin{equation}
    \text{Ca}^* \sim \frac{h}{64 f(\theta) (1+h/a) l},
    \label{eqn:crossover_Ca}
\end{equation}
between flowrate-independent and flowrate-dependent pressure fluctuations [FIG.~\ref{fig:fig_3}(e)]. The above argument suggests two interesting implications. First, one can potentially infer the characteristic pore size of the material from the fluctuations of the pressure signal in \textit{both} viscously-dominated and capillary-dominated flow regimes. This is especially useful when visualization of the flow in pore space is not possible, which is the case in most porous materials. Second, the characteristic $h$, $a$, and $l$ used in this study yield $\text{Ca}^* \approx \frac{10^{-3}}{f(\theta)}$, which reduces to $\text{Ca}^*\;\sim\;10^{-3}$ for drainage and $\text{Ca}^* \sim 10^{-4}$ for imbibition, in agreement with the data in FIG.~3(e). This means that one should expect the transition from capillary-dominated to viscously-dominated flow regimes at different $\text{Ca*}$ in drainage and imbibition. The order of magnitude of $f(\theta)$ was obtained by calculating $\Delta p_\text{crit}$ for all pore throats at $\theta \in \{46^\circ,160^\circ\}$ with the quasi-static model \citep{primkulov18-prf} and taking an average of $f(\theta)=\frac{\gamma}{a \Delta p_\text{crit}}$ for each contact angle. Finally, the viscous pressure fluctuation component scales as $\delta{p_\text{visc}} \sim \mu u$, which is equivalent to $\delta{p_\text{visc}} \sim \text{Ca}$ when interfacial tension is kept constant. This explains the slope of the viscously-dominated portion of the graph in FIG.~\ref{fig:fig_3}(e).

\section{Conclusion}\label{sec:concl}
Overall, our moving-capacitor network model provides new fundamental insights into the dynamics of immiscible fluid-fluid displacement in porous media for a wide range of $\text{Ca}$ and wettabilities. The model completes the picture of the displacement by covering both high and low $\text{Ca}$ which allows, for the first time, to reproduce experimental observations of invading fluid patterns \citep{Zhao2016}, injection pressure and front velocity in drainage \citep{Maloy1992,Furuberg1996,Moebius2012} and imbibition. Our observations and scaling arguments on the transition from viscous-dominated to capillary-dominated flow regime suggest that it is possible to infer the character of the multiphase-flow displacement purely from the injection pressure signal. This poses an exciting prospect for detailed experiments.

\bibliographystyle{jfm}
\bibliography{BibList.bib}

\end{document}